\documentclass[aps,prd,twocolumn,groupedaddress]{revtex4-1}
\usepackage{graphicx}
\usepackage{wrapfig}
\usepackage{latexsym}
\usepackage{amssymb}
\usepackage{amsfonts,color}
\usepackage{float}
\usepackage{amsmath}

\newcommand{\be}{\begin{equation}}
\newcommand{\ee}{\end{equation}}
\newcommand{\ba}{\begin{align}}
\newcommand{\ea}{\end{align}}

\begin{document}


\title{Symmetry breaking, strain solitons and mechanical edge modes in monolayer antimony}
\author{Joshua Chiel$^{(1)}$, Harsh Mathur$^{(1)}$ and Onuttom Narayan$^{(2)}$}
\affiliation{$^{(1)}$Department of Physics, Case Western Reserve University, Cleveland, Ohio 44106-7079}

\vspace{1mm}
\affiliation{$^{(2)}$Department of Physics, University of California, Santa Cruz, CA 95064}


\date{\today}

\begin{abstract}

Two-dimensional materials exhibit a variety of mechanical instabilities
accompanied by spontaneous symmetry breaking. Here we develop a continuum 
description of the buckling instability of antimonene sheets. Regions of 
oppositely directed buckling constitute domains separated by domain walls that
are solitons in our model. Perturbations about equilibrium propagate
as waves with a gapped dispersion in the bulk but there is a gapless
mode with linear dispersion that propagates along the domain walls in a manner
reminiscent of the electronic modes of topological insulators.
We establish that monolayer antimonene is a mechanical topological
insulator by demonstrating a mapping between our continuum 
model and an underlying Dirac equation of the symmetry class BDI
which is known to be a topological insulator in one dimension
and a weak topological insulator in two dimensions. Monolayer
antimony can be produced by exfoliation as well as epitaxy
and the effects predicted in this paper should be accessible
to standard experimental tools such as scanning probe microscopy
and Raman spectroscopy. We surmise that the effects studied
here (namely low scale symmetry breaking, strain solitons and
gapless edge modes) are not limited to antimonene but are 
common features of two dimensional materials. 

\end{abstract}


\maketitle

\section{Introduction \label{introduction}}

Spontaneous symmetry breaking is of fundamental importance
in condensed matter \cite{anderson} and elementary particle physics \cite{peskin}. 
An unusual mechanical example of symmetry breaking was
studied by Alden {\em et al.} in bilayer graphene \cite{alden}. In this material
the two graphene sheets may be stacked on top of one another in 
two distinct stable arrangements called AB and BA stacking.
Domain walls and other more complicated defects were discovered to
form between regions with different kinds of stacking. 
Here we show that an analogous phenomenon occurs in monolayers 
of antimony dubbed antimonene \cite{discovery}. Like graphene, antimonene
has a planar honeycomb structure.
However in antimonene the structure is found to
buckle with the atoms in the A and B sublattices of the honeycomb 
displaced vertically relative to one another \cite{review}. This buckling constitutes a
form of symmetry breaking with the two equivalent equilibria
corresponding respectively to the A sublattice moving upward
or downward relative to the B sublattice. We develop a 
continuum model to describe this symmetry breaking using an
order parameter that corresponds to the (staggered) displacement
of the two sublattices relative to the unstable planar conformation
of the sheet. The model allows us to predict the profile of a
domain wall that would separate regions corresponding to the
two different phases. Moreover within a single phase we find
that waves corresponding to variation in the relative
vertical displacement between the two sublattices have
a gapped dispersion at long wavelength characteristic of
a massive relativistic particle. However we also find a 
gapless linearly dispersing mode that propagates along 
domain walls. This behavior is reminiscent of topological
insulators which have a gapped electronic spectrum in the
bulk but gapless edge states. We demonstrate that our 
mechanical system is in fact a weak topological 
insulator by reformulating the problem in terms of
an underlying Dirac equation. The findings closely
parallel the results of ref \cite{yu} who found that bilayer
graphene was also an example of a mechanical topological
insulator.

Our results are of interest from several points of view.
First, mechanical analogs of topological insulators are
of fundamental interest \cite{kane, lubensky}. While mechanical structures
that are analogous to topological insulators have been
fabricated in the lab \cite{phonon1,phonon2,phonon3,phonon4}, antimonene provides an example
of a natural material that has this behavior. 
Second, the remarkable electronic properties of two
dimensional materials have been a major driver of research
in the field. Notably two dimensional materials are Dirac
metals \cite{geim} and semiconductors \cite{jie}, and are found to exhibit a room
temperature quantum Hall effect \cite{room}, Mott insulating behavior \cite{mott}
and a strongly correlated superconducting state \cite{sc}. Domain
walls in bilayer graphene were predicted to support
electronic zero modes \cite{zhang} that were subsequently observed
experimentally \cite{zhangexp}. One might expect similarly interesting
electronic states to form at domain walls in antimonene.
Finally there have been remarkable advances in the quantum 
control of mechanical resonators and the coupling of these
resonators to superconducting circuits or other qubits. A key challenge
in these experiments is fabrication of a suspended beam
or Bragg mirrors to isolate a surface acoustic wave mode
\cite{schwab}. The spontaneous appearance of
a mechanical zero mode localized along a domain wall may be of 
potential interest from the viewpoint of these experiments.

Antimonene has recently been exfoliated as well as grown 
by epitaxy on a palladium telluride substrate \cite{review}. STM 
measurements show that the atoms are arranged in a 
honeycomb (lattice constant $a = 4.13$\AA) with relative
vertical displacement of the A and B sublattices (estimated
to be $2 \eta = 1.65$\AA) \cite{discovery}. This structure was predicted earlier
based on electronic structure calculations \cite{predict}. Silicene, germanene,
and phosphene also have buckled structures \cite{jie} but here we focus
on antimonene because its structure is particularly simple.
Moreover, antimonene is available both in free standing form
as well as on a substrate, and the substrate interaction does
not appear to significantly modify the structure \cite{review}. We leave open
for future work the study of more complicated buckled 
structures and the rich effects produced by substrates
due to incommensurability. 

In the remainder of this paper we introduce our continuum
model (section II) and use it to determine the profile of a domain
wall (section III). Next we turn to dynamics showing that
in either broken symmetric phase there is a long wavelength mode
that is gapped (section IVa) but that there is a localized mode
with a linear dispersion that propagates along the domain wall
(section IVb). Following Kane and Lubensky \cite{kane}, we develop an 
underlying Dirac description of the zero mode. This analysis
reveals that with regard to this elastic mode antimonene is a
weak topological insulator of the class BDI (section IVc).
We conclude by summarizing our results and open questions (section V). 

\section{The Model}

\label{sec:model}

The order parameter in our model $\zeta$ denotes the vertical
displacement of the A sub-lattice and the negative of the 
vertical displacement of the B sub-lattice. Hence $\zeta = 0$ 
corresponds to a flat sheet of antimonene whereas $\zeta = \pm \eta$
corresponds to the equilibrium separation of $2 \eta$ between
the A and B sublattices. As noted above the measured value of
$2 \eta = 1.65$\AA. Evidently the potential energy density
$V(\zeta)$ has to be a double well symmetric under $\zeta \rightarrow - \zeta$.
The simplest form consistent with this symmetry is 
\begin{equation}
    V ( \zeta ) = \frac{\lambda}{4} ( \zeta^2 - \eta^2 )^2.
    \label{eq:pot}
\end{equation}
Here $\lambda$ is a phenomenological parameter. We see that 
buckling energy per unit cell in the model is 
of order $\lambda \eta^4 a^2$. We expect that this scale
should be of order meV but there is insufficient 
convergence between various first principles calculations 
of the electronic structure of antimonene \cite{review} to reliably
determine $\lambda$ which must therefore be treated as a 
phenomenological parameter. 

Thus far we have assumed a uniform relative displacement
of the two sublattices. We now consider the circumstance
that $\zeta$ varies slowly with position. This would happen for
example if one edge of the sheet has the stable conformation
$\zeta = \eta$ and the other has $\zeta = - \eta$. We expect that
there would be a free energy cost to spatial variation in $\zeta$.
The simplest form consistent with C$_6$ rotational symmetry is
\begin{equation}
    {\cal F} = \int d x d y \; \left[ \frac{1}{2} \sigma ( \nabla \zeta )^2 + V(\zeta) \right].
    \label{eq:free}
\end{equation}
Here $\sigma$ is another phenomenological parameter with units of surface tension
and the integral is over the surface of the antimonene monolayer which is assumed to
lie in the $x$-$y$ plane. We will see below that if domain walls and zero modes are
observed it should be possible to determine the two parameters in our model
$\lambda$ and $\sigma$ from the measured width of the wall and the slope of the linear
dispersion of the zero mode. 

In the following it will be convenient to rescale $\zeta = \eta \overline{\zeta}$,
$(x, y) = (\overline{x},\overline{y}) \ell$ where $\ell = \sqrt{\sigma/\lambda \eta^2}$ and ${\cal F} = \overline{{\cal F}} \lambda \eta^4$.
In terms of the rescaled variables 
\begin{equation}
    {\cal F} = \int dx dy \; \left[ \frac{1}{2} (\nabla \zeta)^2 + \frac{1}{4} ( \zeta^2 - 1 ) \right]
    \label{eq:rescaledfree}
\end{equation}
Note that in eq (\ref{eq:rescaledfree}) and in the following we omit the overlines
although we are working with the rescaled variables. We do so for typographical clarity 
and because it will always be clear from context whether we are working with the original
variables or their rescaled counterparts.

\section{The Wall}

\label{sec:wall} 

In order to model a domain wall we impose $\zeta = \pm 1$ as 
$x \rightarrow \pm \infty$. In equilibrium the displacement field $\zeta$ will minimize the free energy 
(\ref{eq:rescaledfree}) and therefore satisfy
\begin{equation}
    - \nabla^2 \zeta + (\zeta^2 - 1) \zeta = 0. 
    \label{eq:eom}
\end{equation}
Assuming $\zeta$ depends only
on $x$ eq (\ref{eq:eom}) can be integrated to show that
\begin{equation}
    \frac{1}{2} \left( \frac{d \zeta}{d x} \right)^2 - \frac{1}{4} ( \zeta^2 - 1 )^2 
    \label{eq:conserved}
\end{equation}
is a constant independent of $x$. The boundary conditions then reveal that this constant
is zero and eq (\ref{eq:conserved}) may be readily integrated to yield
\begin{equation} 
\zeta = \tanh \left[ \frac{1}{\sqrt{2}} (x - x_0) \right].
\label{eq:profile}
\end{equation} 
Here $x_0$ is a constant of integration that corresponds to the location of the domain wall.
Due to translational symmetry of the model the location of the boundary is arbitrary and we
may choose $x_0 = 0$. 

Reverting to the original variables we may write the profile of the wall as
\begin{equation}
    \zeta_{{\rm W}} = \eta \tanh \left[ \frac{x}{\sqrt{2} \ell} \right]
    \label{eq:naturalprofile}
\end{equation}
where $\ell = \sqrt{\sigma/\lambda \eta^2}$. Eq (\ref{eq:naturalprofile}) is the well-known kink soliton
ubiquitous in theories with a quartic symmetry breaking potential \cite{raja,txv7}. 
We see that the width of the soliton is set by the competition between
the gradient and potential energies with a larger value of the potential
energy favoring a narrower wall. The width should be readily measurable 
using STM and will provide a direct determination of the parameter 
$\ell$. For the quantitative applicability of our
continuum model it is necessary that the width of the wall should be large
on an atomic scale. 

\section{Dynamics}

\label{sec:dynamics}

We can readily incorporate dynamics into our model by recognizing the functional
derivative of the free energy as a restoring force on the antimonene sheet.
The equation of motion is then
\begin{equation}
    \mu \frac{\partial^2}{\partial t^2} \zeta = \sigma \nabla^2 \zeta - \lambda ( \zeta^2 - \eta^2 ) \zeta. 
    \label{eq:wave}
\end{equation}
Here $\mu$ is the mass per unit area of the antimonene sheet ($\mu = 1.6$ mg/m$^2$). 
Introducing a rescaled time $t = \overline{t} \tau$ with $\tau = \sqrt{\mu/\lambda \eta^2}$ 
we can rewrite eq (\ref{eq:wave}) in terms of rescaled variables as
\begin{equation}
    \frac{\partial^2}{\partial t^2} \zeta = \nabla^2 \zeta - (\zeta^2 - 1) \zeta.
    \label{eq:naturalwave}
\end{equation}
Although we are working with rescaled variables in eq (\ref{eq:naturalwave}) we 
omit overlines for clarity as we have done earlier.

\subsection{Bulk behavior}

\label{sec:bulk}

To analyze the behavior of the bulk equilibrium phase we linearize the 
equation of motion (\ref{eq:naturalwave}) by writing $\zeta = \pm 1 + h$ and
expanding to linear order in $h$. We obtain
\begin{equation}
    - \frac{\partial^2 h}{\partial t^2} = - \nabla^2 h + 2 h,
    \label{eq:kleingordon}
\end{equation}
the relativistic Klein-Gordon equation \cite{peskin} which has plane wave solutions
with dispersion 
\begin{equation}
    \omega^2 = k^2 c^2 + \Omega^2
    \label{eq:dispersion}
\end{equation}
where the velocity $c = \sqrt{\sigma/\mu}$ and the gap frequency
$\Omega = \sqrt{2 \lambda \eta^2/\mu}$. Measurement of the dispersion
would therefore allow separate determination of the model parameters
$\lambda$ and $\sigma$; the additional measurement of the domain wall
width would provide a separate non-trivial check. Physically $k \rightarrow 0$
corresponds to uniformly varying the staggered displacement $\zeta$ away from
its equilibrium value; the finite gap frequency $\Omega$ simply reflects the
existence of a restoring force towards equilibrium.
Parenthetically we note that there is a second mode of vertical oscillation
of antimonene in which both sublattices are displaced in the same 
direction. This mode is gapless and linearly dispersing in the $k \rightarrow 0$
limit. 

\subsection{Gapless Boundary Mode}

\label{sec:gapless}

Now let us analyze the modes of an antimonene sheet with a domain wall present.
To this end we write $\zeta = \zeta_{{\rm W}} (x) + h(x,y,t)$ where $\zeta_{{\rm W}}$
is the domain wall profile (\ref{eq:naturalprofile}). To linear order in $h$ the
equation of motion is found to be 
\begin{equation}
    - \frac{\partial^2}{\partial t^2} h = - \nabla^2 h + (3 \zeta_{{\rm W}}^2 - 1 ) h.
    \label{eq:wallstab}
\end{equation}
We look for solutions propagating along the domain wall of the form 
$h = \psi (x) \exp( i k y - i \omega t)$. The form factor $\psi$ then obeys
a one dimensional Schr\"{o}dinger like equation
\begin{equation}
    - \frac{d^2}{d x^2} \psi + U(x) \psi = (\omega^2 - k^2) \psi
    \label{eq:schrodinger}
\end{equation}
Here the effective potential $U(x) = 3 \zeta_{{\rm W}}^2 - 1$ is a symmetric well
with a minimum value of $-1$ for $x = 0$ and asymptotic behavior $U \rightarrow 2$ 
as $x \rightarrow \pm \infty$. We focus on bound states for which $-1 < \omega^2 - k^2 < 2$;
these states correspond to propagating modes that are localized near the wall. 

It is easy to verify that
\begin{equation}
    \psi_{{\rm zero}} (x) = \frac{d \zeta_{{\rm W}} }{d x} = \frac{1}{\sqrt{2}} {\rm sech}^2 \left( \frac{x}{\sqrt{2}} \right)
\end{equation}
is a solution to eq (\ref{eq:schrodinger}) with eigenvalue $\omega^2 - k^2 = 0$. This solution is called the zero mode
and it corresponds to a gapless mode propagating along the domain wall with a linear dispersion $\omega^2 = k^2$. 
The existence of this solution is traceable to the translational invariance of the problem, namely, the freedom to
choose the location of the domain wall. 

Having obtained the zero mode we can now use an argument standard in supersymmetric quantum mechanics
to gain further insight into the solutions to eq (\ref{eq:schrodinger}). The existence of the zero 
mode implies the factorization
\begin{equation}
- \frac{d^2}{dx^2} + U(x) = A^\dagger A
\label{eq:factor}
\end{equation}
where
\begin{equation}
    A = \frac{d}{dx} - \frac{d \ln \psi_{{\rm zero}}}{dx} = \frac{d}{dx} + \sqrt{2} \tanh \left[ \frac{x}{\sqrt{2}} \right]
    \label{eq:a}
\end{equation}
The form $A^\dagger A$ reveals that the eigenvalues must be non-negative
demonstrating that the zero-mode is the bound mode with lowest eigenvalue. 
We expect eq (\ref{eq:schrodinger}) to have at least one bound mode because
it is a one dimensional potential well but using the factorized form 
we can now demonstrate that it has at least one other bound mode. 
This is because the operator
\begin{equation}
    A A^\dagger = - \frac{d^2}{d x^2} + W(x) 
\label{eq:susy}
\end{equation}
has the same spectrum as $A^\dagger A$ except for the zero mode.
Explicitly we find $W(x) = 1 + \tanh^2 (x/\sqrt{2})$ which is also symmetric well
(albeit as shallower one than $U$). 
The well has minimum value $W =1$ at the origin and asymptotic behavior $W \rightarrow 2$
as $x \rightarrow \pm 1$. Hence the operator $A A^\dagger$ also 
must also have at least one bound mode.
The bound modes of $A A^\dagger$ represent additional bound modes of $A^\dagger A$,
the operator in eq (\ref{eq:schrodinger}). 

We have presented the supersymmetry argument above because of its generality.
However for the particular form of $U$ for the kink soliton it is well known
that eq (\ref{eq:schrodinger}) is exactly soluble \cite{raja, txv7}. The exact analysis reveals
that in addition to the zero mode there is exactly one other bound mode
$\psi_{{\rm ex}} = \tanh (x/\sqrt{2}) \; {\rm sech} (x/\sqrt{2})$ with eigenvalue
$3/2$ corresponding to the dispersion
\begin{equation}
    \omega_{{\rm ex}}^2 = \frac{3}{2} + k^2.
    \label{eq:breath}
\end{equation}
Physically the zero mode for small $k$ corresponds to a slow undulation in the displacement
of the wall perpendicular to itself. The excited bound mode corresponds to slow oscillations 
in the width of the wall.

Although we are primarily interested in bound modes localized near the domain wall
for the record we note that there are also unbound solutions to eq (\ref{eq:schrodinger}).
These modes correspond to the gapped bulk modes analyzed in the preceding section 
scattering off the domain wall. Far from the domain wall these modes will have the
form of plane waves $\psi (x) \rightarrow \exp( i p x )$. The exact solution is found to be \cite{raja, txv7}
\begin{equation}
    \psi (x) = e^{i p x} \left[ 3 + \tanh^2 \frac{x}{\sqrt{2} } - 1 - 2 p^2 
    - 3 \sqrt{2} i p \tanh \frac{x}{\sqrt{2}}   \right]
    \label{eq:scatter}
\end{equation}
This has the asymptotic behavior
\begin{equation}
    \psi (x) \rightarrow \exp \left[ i p x \mp \frac{1}{2} \delta (p) \right] \hspace{3mm} {\rm as} 
    \hspace{3mm} x \rightarrow \pm \infty 
    \label{eq:phaseshift}
\end{equation}
where the phase shift $\delta(p) = 2 \tan^{-1} [ 3\sqrt{2} p/(2 - 2 p^2) ]$. 
This solution reveals the unusual feature of the potential $U$ that it is reflectionless.
An incoming wave from the left is transmitted with a phase shift but without any reflection. 
These unbound modes have dispersion
\begin{equation}
    \omega^2 = (k^2 + p^2) c^2 + \Omega^2
\end{equation}
consistent with the result for bulk modes derived in the previous section. 

\subsection{Topological Considerations}

\label{sec:topology}

The existence of gapped bulk modes and a gapless edge mode is reminiscent of the behavior
of electronic states in a topological insulator. It is therefore tempting to look for a 
topological interpretation of the edge modes analyzed above. There is at present no 
completely general theory of topological modes in mechanical systems. However for a restricted
class of lattices of mass points connected by springs Kane and Lubensky \cite{kane} have developed
a topological analysis. Their key observation is that for a system of mass points connected by springs one can write
the equation of motion in the form 
\begin{equation}
    \frac{d^2}{dt^2} \xi = - Q^T Q \xi. 
    \label{eq:kanelubensky}
\end{equation}
Here $\xi$ is a column vector the entries of which are the components of the displacement from equilibrium
of each mass point. The matrix $Q$ relates these displacements to the extensions of the springs (for simplicity
we have assumed that the mass points all have the same mass $m$ and the springs the same spring constant $k$
and we have set $m = k =1$). The matrix $Q$ will in general be rectangular unless the number of springs $M$
equals $N$, the number of components of the displacement vector. The topological analysis of Kane and Lubensky
is restricted to the isostatic case $M = N$.

To carry out a normal mode analysis of eq (\ref{eq:kanelubensky}) one should determine the eigenvalues of $Q^TQ$ which
correspond to the squares of the normal mode frequencies. 
Kane and Lubensky however construct the ``Hamiltonian''
\begin{equation}
    {\cal H} = \left( 
    \begin{array}{cc}
    0 & Q^T \\
    Q & 0 
    \end{array}
    \right) 
        \label{eq:ham}
\end{equation}
The eigenvalues of ${\cal H}$ are either zero or come in pairs with opposite sign due to the
particle hole symmetry of the Hamiltonian, 
$[ C, {\cal H} ]_+ = 0$. Here $C$ is the anti-unitary operator
\begin{equation}
    C \psi = \left(
    \begin{array}{cc}
    1 & 0 \\
    0 & -1 
    \end{array} \right) 
    \psi^\ast 
    \label{eq:conjugation}
\end{equation}
The squares of the eigenvalues of ${\cal H}$ match the eigenvalues of $Q^TQ$ establishing
a mapping between the Hamiltonian and the normal mode problem. The topological analysis 
is then carried out on the Hamiltonian ${\cal H}$. 
Since ${\cal H}$ also has time reversal symmetry $[ T, {\cal H}] = 0$ (where the time reversal operator
$T$ consists of complex conjugation), and since $C^2 = 1$ and $T^2 =1$, this Hamiltonian
belongs to the symmetry class BDI \cite{tenfold}. Hamiltonians in this class are well-known to be topological
insulators in one dimension and to be weak topological insulators in two dimensions \cite{tenfold}.

Here we work with a continuum description of the mechanical problem so we cannot directly
apply the above analysis. However following their general strategy we find a ``square-root''
of the continuum mechanical model in terms of a Dirac equation. This mapping allows
us to draw upon a comprehensive topological analysis of Dirac equations \cite{tenfold} as shown
below. 

\subsubsection{One dimension}

We consider a one dimensional analog of our antimonene model which is governed by the exact equation of motion 
\begin{equation}
    - \frac{\partial^2}{\partial t^2} \zeta = - \frac{\partial^2}{\partial x^2} \zeta + V'(\zeta).
    \label{eq:onedim}
\end{equation}
For simplicity, we assume that $V$ is a symmetric function and that it has two minima
$V (\pm \eta) = 0$. Other than that we impose no restriction on $V$. In particular, it is not
necessarily of the quartic form eq (\ref{eq:pot}).
A domain wall $\zeta_W(x)$ is a solution to the static equation
\begin{equation}
    - \frac{d^2}{d x^2} \zeta_W + V'(\zeta_W) = 0
    \label{eq:domaingen1d}
\end{equation}
subject to the boundary condition $\zeta_W \rightarrow \pm \eta$ as $x \rightarrow \pm \infty$. 
Linearizing the equation of motion (\ref{eq:onedim}) about the soliton solution $\zeta_W$ 
we are led to the Schr\"{o}dinger operator
\begin{equation}
    - \frac{d^2}{d x^2} + U (x) \hspace{3mm} {\rm with} \hspace{3mm} U(x) = V''[\zeta_W(x)].
\label{eq:schrodingerop}
\end{equation}
By differentiating eq (\ref{eq:domaingen1d}) it is easy to verify that 
\begin{equation}
    \psi_g (x) = \frac{d}{dx} \zeta_W(x) 
    \label{eq:zeroschrodinger}
\end{equation}
is a zero mode of the Schr\"{o}dinger operator (\ref{eq:schrodingerop}).
We see that the zero mode is extremely robust and exists independently of the
form of $V(\zeta)$. 
Next we observe
\begin{equation}
    U(x) = \frac{1}{\psi_g} \frac{d^2}{d x^2} \psi_g.
    \label{eq:anotheru}
\end{equation}
Defining
\begin{equation}
    A = - \frac{d}{dx} + \frac{1}{\psi_g} \frac{d}{dx} \psi_g 
    \label{eq:agen1d}
\end{equation}
it is now easy to verify that the Schr\"{o}dinger operator (\ref{eq:schrodingerop}) is
equal to $A^\dagger A$. Note that this factorization is also extremely general. It does
not depend on the specific form of $V(\zeta)$; in particular the standard quartic form is
not essential. 

Thus far we have simply introduced a one dimensional version of our model of antimonene
and recounted well-known results about its soliton and zero mode solutions \cite{raja,txv7}. Now, following
Kane and Lubensky, we show that this model has a remarkable mapping to an underlying Dirac
equation which coincides with the Jackiw-Rebbi model \cite{jackiw}. 
By analogy to the mapping from the
normal mode operator $Q^T Q$ to the Hamiltonian eq (\ref{eq:ham}) we go from the
Schr\"{o}dinger operator $A^\dagger A$ to the underlying Hamiltonian
\begin{equation}
    {\cal H}_D = \left(
    \begin{array}{cc}
    0 & A^{\dagger} \\
    A & 0 
    \end{array}
    \right)
    \label{eq:ham1d}
\end{equation}
Comparing eqs (\ref{eq:agen1d}) and (\ref{eq:ham1d}) to the one-dimensional Dirac Hamiltonian
\begin{equation}
    {\cal H}_D = - i \alpha \frac{\partial}{\partial x} + m (x) \beta
    \label{eq:diracstandard1d}
\end{equation}
we get a match if we adopt the representation $\alpha = - \sigma_y, \beta = - \sigma_x$ and
$m = - (1/\psi_g)(d \psi_g/d x)$. The Dirac Hamiltonian commutes with time reversal (defined as the
operation ${\cal T} \psi = \psi^\ast$) and anticommutes with charge conjugation (defined as the
operation ${\cal C} \psi = \sigma_z \psi^\ast$). Since ${\cal T}^2 = 1$ and ${\cal C}^2 = 1$,
it follows that the Dirac Hamiltonian belongs to the symmetry class BDI. 
Note moreover that the mass satisfies 
$m \rightarrow \pm m_0$ for $x \rightarrow \pm \infty$. Here $m_0 = \sqrt{V''(\eta)}$. This follows 
from the asymptotic behavior of $\zeta_W$ as $x \rightarrow \pm \infty$ which can be established
from eq (\ref{eq:domaingen1d}). 
Thus we see that the Dirac equation underlying our mechanical model has a mass that
varies with position and asymptotically approaches values with opposite sign as
$x \rightarrow \pm \infty$. This is precisely the Jackiw-Rebbi model which is well
known to have a zero mode of its own \cite{raja,txv7,jackiw}. The zero mode can be constructed explicitly.
Up to a normalization constant it is given by
\begin{equation}
    \psi_{{\rm zero}} \propto 
    \left( \begin{array}{c}
    \exp [ - M(x) ] \\
        0 \end{array} \right) 
    \label{eq:diraczero1d}
\end{equation}
where $M(x) = \int_0^x d x' \; m(x')$. 
The existence of the zero mode can also be assured by calculation of a topological invariant,
the winding number \cite{tenfold}. 

In summary we have established that the one dimensional counterpart of our
antimonene model has a domain wall and a zero mode localized at the wall. 
Moreover we have established a mapping of our model to an underlying Dirac equation,
the Jackiw-Rebbi model. This mapping shows that the zero mode of our mechanical
model has a topological origin. 

\subsubsection{Two dimensions}

Our objective in this section is to find a Dirac equation that underlies our 
continuum mechanical model of an antimonene sheet. In the bulk this is 
simply the Klein-Gordon equation (\ref{eq:kleingordon}).
If we write the Hamiltonian as 
\begin{equation}
    {\cal H}_D = - i {\boldsymbol \alpha} \cdot \nabla + \beta m
    \label{eq:dirac2d}
\end{equation}
then upon squaring the Hamiltonian we are sure to obtain the 
Klein Gordon equation provided the matrices ${\boldsymbol \alpha}$ and $\beta$
obey the Dirac algebra and $m$ is a constant. In two dimensions the simplest choices
for the Dirac matrices are $\alpha_1 = \sigma_1$, $\alpha_2 = \sigma_2$ and $\beta = \pm \sigma_3$;
all other $2\times 2$ representations are unitarily equivalent to one of these choices. 
However with this choice the Dirac Hamiltonian lacks time reversal symmetry and does
not have the off diagonal structure of eq (\ref{eq:ham}); hence we are compelled to use
a $4\times4$ representation. In the following we will find it convenient to adopt
\begin{equation}
    \alpha_1 = \left[
    \begin{array}{cc}
    0 & - i \sigma_1 \\
    i \sigma_1 & 0
    \end{array} \right], 
    \alpha_2 = \left[
    \begin{array}{cc}
    0 & - i \sigma_3 \\
    i \sigma_3 & 0 
    \end{array} \right], 
    \beta = 
    \left[
    \begin{array}{cc}
    0 & - i \sigma_2 \\
    i \sigma_2 & 0 
    \end{array} \right].
    \label{eq:diracrep}    
\end{equation}
This Hamiltonian has the symmetries $[{\cal H}_D, {\cal T} ] = 0$ and 
$[ {\cal H}_D, {\cal C} ]_+ = 0$ where time reversal ${\cal T} \psi = \psi^\ast$
and conjugation is the antiunitary symmetry
\begin{equation}
    {\cal C} \psi = \left(
    \begin{array}{cc}
    1 & 0 \\
    0 & - 1
    \end{array}
    \right) \psi^\ast.
    \label{eq:conjugate2d}
\end{equation}
Note that ${\cal T}^2 = 1$ and ${\cal C}^2 = 1$; hence the Hamiltonian belongs to the
intended class BDI.

If we now allow for the possibility that
$m$ depends on $x$ (but not on $y$) we obtain
\begin{equation}
    {\cal H}_D^2 = - \nabla^2 + m^2 - i \alpha_1 \beta \frac{\partial m}{\partial x} 
    \label{eq:diracsquare2d}
\end{equation}
For the representation (\ref{eq:conjugate2d}) 
\begin{equation}
    - i \alpha_1 \beta = \left(
    \begin{array}{cc}
    \sigma_3 & 0 \\
    0 & \sigma_3
    \end{array}
    \right);
    \label{eq:alphabeta}
\end{equation}
Hence ${\cal H_D}^2$ corresponds to the two Schr\"{o}dinger operators
\begin{equation}
    {\cal H}_{{\rm S}}^\pm = - \nabla^2 + U_\pm (x) \hspace{3mm} {\rm with} \hspace{3mm} U_\pm (x) = m^2 \pm \frac{\partial m}{\partial x}.
    \label{eq:schrodingerpm}
\end{equation}
These operators are supersymmetric partners in the sense that they have the same spectrum except possibly for
zero modes. 

Having constructed the underlying Dirac model we now turn to its zero modes. 
The zero modes of ${\cal H}_D$ can be explicitly constructed. We assume
that $m(x)$ changes sign along the $y$-axis which is the location of a domain
wall and that $m(x) \rightarrow \pm m_0$
as $x \rightarrow \infty$. We seek solutions to the eigenvalue equation
${\cal H}_D \psi = \omega \psi$ that propagate along the $y$-direction 
and are localized along the wall. We find two chiral modes
\begin{equation}
    \psi_\pm^{{\rm zero}} \propto
    \left[ \begin{array}{c}
    0 \\
    1 \\
    0 \\
    \pm i \end{array}
    \right] 
    \exp[ - M(x) ] \exp(i k y)
    \label{eq:dirac2dzeromodes}
\end{equation}
with linear dispersion $\omega = \pm k$. These modes are chiral in the
sense that the plus mode propagates along the positive $y$ axis while
the minus mode propagates along the negative $y$ axis. These modes
are robust in the sense that they exist independent of the precise form
of $m(x)$. However the chiral modes are protected from backscattering
only by translational invariance in the $y$-direction. If the domain wall meanders (or equivalently
$m$ depends on $y$ as well as $x$) then the modes are no longer protected from
backscattering---or even guaranteed to exist. In this sense the BDI Dirac equation 
is a weak topological insulator in two dimensions. 

For a suitable choice of $m(x)$ we can arrange for the Schr\"{o}dinger operator 
${\cal H}_{{\rm S}}^-$ to match eq (\ref{eq:schrodinger}) which describes
the modes of our continuum mechanical model in the presence of a domain wall.
The existence of zero modes in the continuum mechanical model that are localized
on the domain wall can then be traced back to the chiral modes of the underlying
Dirac equation. The robustness of the Dirac modes to variations in $m(x)$ corresponds
to a robustness in the modes of the mechanical model under variations in the form
of the symmetry breaking potential (eq \ref{eq:pot}). 

\section{Summary and Conclusion}

\label{sec:conclusion} 

In this paper we surmise that mechanical symmetry breaking will prove ubiquitous in 
two dimensional crystals due to competing stacking conformations, buckling instabilities
and interaction with substrates, resulting in defects at the boundaries of stable
domains of different symmetry. Because of the low energy scale involved these defects
may be described in terms of continuum theories. As a concrete example of these
general considerations here we examine in detail the case of monolayer antimonene. 

A horizontal planar honeycomb 
arrangement of the antimony atoms is unstable to buckling wherein the $A$ and $B$ 
sublattices of the honeycomb are separated vertically. There are two equivalent
stable configurations depending on whether the $A$ sublattice moves upward or
downward relative to the $B$ sublattice. To describe this instability we develop
a continuum model in which the degree of buckling is the order parameter. 
This model reveals that the domains of positive and negative
buckling are separated by walls whose profile and width we can precisely calculate
in terms of the model parameters. We find that perturbations about the equilibrium
propagate as waves with a gapped dispersion in a pure domain but along the domain
walls there is an additional gapless mode. Physically this mode corresponds to 
undulations in the position of the domain wall and is due to the underlying
translational invariance of the continuum model. This behavior of gapped
modes in the bulk and gapless modes on the boundary is reminiscent of topological
insulators. Following Kane and Lubensky we map our continuum model to an underlying
Dirac equation. This mapping allows us to identify monolayer antimonene as a 
mechanical topological insulator, albeit a weak one. 

Among open problems for future work we mention three. First, it would be desirable
to experimentally observe the effects discussed in the paper. These should be 
amenable to standard tools like scanning probe microscopy and Raman spectroscopy.
Second, it would be of interest to study electronic defect states such as electronic
zero modes that might potentially propagate along domain walls. Finally, there is an
abundance of two dimensional materials with similar low energy mechanical symmetry 
breaking and defects that deserve further exploration.

\section{Acknowledgement}

\label{sec:ack}

We are grateful to Walter Lambrecht and Santosh Radha for telling us about
the buckling instability of antimonene and bringing ref \cite{review} to our 
attention. We thank Michael Lawler for discussions and bringing ref \cite{kane}  
to our attention.

\end{document}